
\magnification 1200
\hsize=31pc
\vsize=55 truepc
\baselineskip=26 truept
\hfuzz=2pt
\vfuzz=4pt
\pretolerance=5000
\tolerance=5000
\parskip=0pt plus 1pt
\parindent=16pt
\def\avd#1{\overline{#1}}
\def\avt#1{\langle#1\rangle}

\def\s{\sigma}
\def\S{{\bf s}}
\def\Sz{{\bf s}_0}
\def\sq{{\bf s_{(4)}} }
\def\sc{{\bf s_{(5)}} }
\def\P {{\cal{ P}}}
\def\Z {{\cal{ Z}}}
\def\U {{\cal{ U}}}
\def\J {{\widetilde{ J}}}
\def\sp {{\widetilde{ \sigma}}}
\def\x {{\widetilde{ x}}}
\def\b {{\widetilde{ \beta}}}
\def\f {{\widetilde{ f}}}
\vskip 1.truecm\noindent
\centerline{\bf Bethe-Peierls approximation for  the 2d random Ising model}
\vskip 1.4truecm\noindent
\centerline{ G. Paladin$^{1,2}$ and M. Serva$^{3,4}$}
\vskip .4truecm
\centerline{\it $^{1}$
NORDITA, Blegdamsvej 17,  Copenhagen  \O $ \ $DK-2100 Denmark}
\vskip .7truecm
\centerline{\it $^{2}$
Permanent address: Dipartimento di Fisica,  Universit\`a dell'Aquila}
\centerline{\it I-67100 Coppito, L'Aquila, Italy}
\vskip .7truecm
\centerline{$^{3}$ \it Institut Superieure Polytechnique
 de Madagascar}
\centerline{ \it Lot II H 31 Ter, Ankadindramamy
101 Antananarivo, Madagascar}
\vskip .7truecm
\centerline{\it $^{4}$Permanent address:
 Dipartimento di Matematica,  Universit\`a dell'Aquila}
\centerline{\it I-67100 Coppito, L'Aquila, Italy}
\vskip 1.6truecm
\centerline{ABSTRACT}
\vskip .4truecm
The  partition function of the $2d$ Ising model with
random nearest neighbor  coupling  is expressed
 in the dual lattice made of square plaquettes.
The dual model is solved in the the mean field and in different
types of Bethe-Peierls approximations, using the replica method.
\hfill\break
\hfill\break
\noindent
PACS NUMBERS: 05.50.+q, 02.50.+s
\vfill\eject
\vskip .4truecm
{\bf INTRODUCTION}
\medskip

The application of methods of mean field type to Ising models
 allows one to obtain  very accurate approximations of the thermodynamic
quantities.
 However, in presence of quenched disorder this approach is of difficult
implementation.
 In this paper, we show that rather good results can be obtained
 after performing  a duality transformation
of the Ising model
 with random nearest neighbor coupling that assume
 the values $J_{ij}=\pm1$ with equal probability.
The model is thus defined on a dual lattice where the spin
variables are attached to the square plaquettes.
The advantage is that the quadratic term of the dual Hamiltonian
has constant coefficients instead of random ones.
It is therefore possible to use the standard methods of the
 mean field to estimate the quenched free energy.
 Our results can be generalized to higher dimensions, although the
approximations become more rough, because the number of plaquette spins
 over the number of interaction links increases with the dimensionality
 [1]. In particular we obtain an extremely good estimate of the ground
state energy of the random Ising model, by applying the Bethe-Peierls
 approximation where  part of the short range order is taken into account.

In section 1,  we introduce the dual lattice made of elementary square
plaquettes. On this lattice the partition function
  can be expressed as a function  of the inverse temperature
$\b=-{1\over 2} \ln \tanh(1/T)$ where $T=\beta^{-1}$
 is the temperature of the original lattice.

In section 2, we apply the
 mean field approximation to the dual model, using the replica method.

In section 3, we introduce the Bethe-Peierls approximation.
This allows us to obtain a very precise estimate of the ground state energy
of the two dimensional Ising model with random coupling.

In section 4, we show that it is possible to improve the
Bethe-Peierls approximation by introducing an interaction between
different replicas.

In section 5,  the reader will find  some remarks and conclusions.
\bigskip
{\bf 1. DUALITY TRANSFORMATION}
\medskip
\noindent
The partition function
 of the  Ising models
 on a  lattice of $N$ sites
 with nearest neighbor couplings $J_{ij}$ which are
independent identically distributed random variables,
 in absence of external magnetic field, is
$$
Z_N(\beta, \{J_{ij}\})  =  \sum_{ \{ \sigma \} }  \prod_{(i,j)}
\exp (\beta J_{ij} \sigma_i \sigma_j )
\eqno(1)
$$
where the sum runs over the $2^N$ spin configurations $\{\s\}$, and
 the product over  the $2N$  nearest neighbor sites $(i,j)$.
One is interested in computing the quenched free energy
$$
f=- \lim_{N \to \infty} {1\over \beta N} \avd{\ln Z_N}
\eqno(2)
$$
where $\avd{A}$ indicates the average of an observable $A$
 over the distribution of the random coupling.
The quenched free energy is a self-averaging quantity,
 i.e. it is obtained in the thermodynamic limit
 for almost all realizations of disorder [2].

  On the other hand, it is trivial to compute
the so-called annealed free energy
$$
f_a=-\lim_{N \to \infty} {1\over \beta N} \ln \avd{ Z} \ ,
\eqno(3)
$$
corresponding to the free energy of a system where the random coupling
 are not quenched but can thermalize with a relaxation time comparable
 to that one of the spin variables.
In our model, where the coupling are independent
 dichotomic random variables $J_{ij}=\pm 1$ with equal probability,  one has
$$
f_a= - \beta^{-1} \, \ln (2  \cosh^2 \beta \, )
\eqno(4)
$$
However, $f_a$ is a very poor approximation of the quenched free energy,
 and is not able to capture the qualitative features of the model.

In order to estimate (1),
 it is convenient to use the link variable $x_{ij}=\s_i \, \s_j$,
since only terms corresponding to products of the variables $x_{ij}$
 on close loops survive after summing over the spin configurations:
 on every close loop of the lattice $\prod x_{ij}=1$,  while
 $\prod x_{ij}=\s_a \, \s_b$ for a path from the site $a$ to the site $b$.
   A moment of reflection shows that it is sufficient to fix
 $\prod x_{ij}=1$ on the
 elementary square plaquettes $\P$
 to automatically fix  it on all the close loops.
The partition function thus becomes
$$
Z_N(\beta, \{J_{ij}\})  =  \sum_{ \{ x_{ij} \} }   \
\prod_{i=1}^{N_p}  {1+\x_i\over 2} \, \prod_{(i,j)}  e^{\beta J_{ij} x_{ij}}
\eqno(5)
$$
where the number of plaquettes is
$$N_p=N$$
 and  we have introduced the plaquette variable
$\x_i=\prod_{\P_i} x_{ij}$.

For dichotomic random coupling $J_{ij}=\pm 1$ with equal probability,
 the free energy of the model is invariant under the gauge transformation
$x_{ij} \to J_{ij} \, x_{ij}$, so that one has
$$
Z_N  =  \sum_{ \{ x_{ij} \} }   \
\prod_{i=1}^{N}  {1+\x_i \J_i \over 2} \, \prod_{(i,j)}  e^{\beta x_{ij}}
\eqno(6)
$$
where $\J_i=\prod_{\P_i} J_{ij}$ is again a dichotomic random variable
 (the `frustration' [3] of the plaquette $\P_i$).
It is worth remarking that (6) gives the partition function in terms of a sum
 over the $2^{2N}$  configurations of the independent random variables
 $x_{ij}=\pm 1 $   with probability
$$
p_{ij}= {e^{\beta x_{ij}} \over 2  \cosh  \beta}
\eqno(7)
$$
 In this section we shall indicate
the average of an observable $A$
 over such a normalized weight  by
$$
\avt{A} \equiv \sum_{ \{x_{ij} \}  } \prod_{(i,j)} p_{ij} \,  A \, ,
$$
e.g.  one has
 $\avt{x_{ij} }=\tanh \beta$ and $\avt{\x_i}=\tanh^4 \beta$.
With such a notation, the partition function assumes the compact form
$$
Z_N= 2^{N} \, \cosh^{2 N} (\beta)
  \ \avt{ \, \prod_{i=1}^{N}(1 + \x_i \J_i) \, }
\eqno(8)
$$
In order to estimate the average in (8),
let us introduce the dual lattice [4]
 where the sites are located at the centers of each square
  of the original lattice.
A dual spin variable is attached to each
 square plaquette and can assume only the values $\sp_i=\pm 1$
 with equal probability,
 so that one has the identity
$$
1 + \x_i \J_i \,  = \sum_{\sp_i=\pm1} (\x_{i}\, \J_i)^{(1+\sp_i)/2}.
\eqno(9)
$$
Since there is a one-to-one correspondence between links on the original and
on the dual lattice, we can estimate the link-average noting that
$$
\avt{ \prod_{i=1}^{N} \x_i^{{(1+\sp_i)\over 2} }} =
   \avt{ \prod_{(i,j)}  x_{ij}^{{(1+\sp_i)\over 2}
 +{(1+\sp_j)\over 2} } } =
\prod_{(i,j)} \left( \tanh\beta \right)^{ {1-\sp_i \sp_j\over 2 } }
\eqno(10)
$$
The last equality in (10)
 follows from the fact that
$$ x_{ij}^{ {(1+\sp_i)\over 2} +{(1+\sp_j)\over 2}} =
\eqalign{ <x_{ij}> = \tanh\beta
 \qquad {\rm if } \ \sp_i\neq \sp_j\cr
1         \    \qquad {\rm if } \  \sp_i=  \sp_j}
\eqno(11)
$$
 Inserting (10) and (9) into (8) one has
$$
Z_N=2^{N} \, \cosh^{2 N} (\beta) \ e^{-N \, 2 \b } \,
 \sum_{\{ \sp \}} \ \prod_{i=1}^{N} \J_i^{(1+\sp_i)/2}
\, \prod_{(i,j)} e^{\b \, \sp_i \sp_j }
\eqno(12)
$$
where we have introduced the variable
 $$
\b= - \, {1\over 2} \, \ln \, \tanh \, \beta
\eqno(13)
$$
which is the inverse temperature of the dual model
 vanishing as $e^{-2\beta}$
 when the temperature $T=\beta^{-1} \to 0$.
The quenched free energy (2) thus becomes
$$
-\beta \, f(\beta)=  \ln \sinh(2\beta)  -  \, \b \, \f(\b)
\eqno(14)
$$
where $\f$ is the free energy of the dual model, defined as
$$
\f(\b)= - \lim_{N \to \infty}
 {1\over\b  N} \ln \Z_{N}
\eqno (15)
$$
in terms of the partition function
$$
\Z_{N}=\sum_{\{\sp_i\}} e^{\b \sum_{(i,j)}
 \sp_i \, \sp_j}
 \ \prod_{i=1}^{N} \J_i^{(1+\sp_i)/2}
\eqno(16)
 $$
 From (16) the Hamiltonian of the dual model can be defined via
 the relation $\Z_{N_p}=\sum_{\{\sp\}}  e^{-\b H}$,  as
$$
H=
  - \sum_{(i,j)}\sp_i \sp_j \,   - \,  \sum_{i=1}^{N_p} \, \ln(\J_i)
 \,  {(1+\sp_i)\over 2 \b}
\eqno(17)
$$
Let us stress that the quadratic term of the dual Hamiltonian
is independent of the random coupling and
 the randomness enters via a random complex magnetic field
 that can assume the two values $0$ and $i\pi/(2\b)$ with equal probability.
 In fact,
 the weight $\exp(-\b \, H)$  does not define
 a standard Gibbs probability measure on the dual lattice:
  it defines  a signed probability measure,
 differing from that one of the pure Ising model only
 for the presence of the random sign
related to the frustrations of the square plaquettes $\{\J_i\}$.
\bigskip
{\bf 2. REPLICA TRICK AND MEAN FIELD APPROXIMATION}
\medskip
\noindent
The introduction of the dual model allows one
 to apply the mean field approximation,
 since one can easily linearize the Hamiltonian (17) by
neglecting fluctuations.
 A similar method  has been introduced in the framework of field theory
 in statistical systems without disorder such as lattice gauge theories or
 spin models [5-6].

For our purposes, it is convenient to use the replica method
in order to get the quenched free energy of the dual model as
$$
\f(\b)=- \lim_{n \to 0} \lim_{N \to \infty}
{1\over \b n N} \ln \avd{(\Z_N)^n}
\eqno(18)
$$
Let us thus consider $n$ non-interacting replicas of our disordered
 system labelled by $\alpha=1,\cdots,n$.
 Now,  the $\J$'s are
are  independent random variables in $2d$.
 Indeed, one can easily verify that
 $\avd{\prod_i \J_i}=\prod_i \avd{J_i}$
 because $\J_i=\pm 1$ with equal probability.
It is worth noting that this is not true in $3d$,
 where the  $\J_i$ of the $i$ square plaquette of a cube
 can be obtained as a product of the remaining
five $\J$'s of the cube, implying $\prod_{cube} \J_k \equiv 1$,
 so that $\avd{\prod_{cube} \J_k}= 1$ while $(\, \avd{\J_k} \, )^6=0$.
 On the contrary,  a plaquette frustration $\J_i$
  cannot be expressed as a product of the
other ones  in $2d$.
 As a consequence, from (16)
 the partition function of $n$ replicas becomes
$$
\avd{ (\Z_N)^n}=
       \sum_{\{\S\}}
e^{\b \sum_{\alpha=1}^n \sum_{(i,j)}
 \sp_i^{(\alpha)} \, \sp_j^{(\alpha)}}
 \ \prod_{i=1}^{N}  \, \avd{ \prod_{\alpha=1}^n
\J_i^{{(1+\sp_i^{(\alpha)})/ 2}} \,  }
\eqno(19a)
$$
where the sum in (19a) runs over the $2^{Nn}$ spin configurations $\{\S\}$
of the replicas, and we use the compact notation:
$$
\{\S\}   \equiv \{\sp^{(1)}\},\cdots,\{\sp^{(n)}\}
$$
One can easily perform the disorder average in (19a) and gets
$$
\avd{ (\Z_N)^n}      =
      \sum_{\{\S\}}
e^{\b \sum_{\alpha} \sum_{ (i,j)}
 \sp_i^{(\alpha)} \, \sp_j^{(\alpha)} }
 \ \prod_{i=1}^{N}  \, {1  \over 2 }
 \left(
  1+ (-1)^n \prod_{\alpha} \sp_i^{(\alpha)}
 \right)
\eqno(19b)
$$
As the free energy is invariant under the
gauge transformation $\sp_i^{(\alpha)} \to -\sp_i^{(\alpha)}$,
 (19b) assumes the simpler form
$$
\avd{(\Z_N)^n}=
       \sum_{\{\S\}}
e^{\b \sum_{\alpha} \sum_{(i,j)}
 \sp_i^{(\alpha)} \, \sp_j^{(\alpha)}}
 \ \prod_{i=1}^{N}  \, {1    \over 2}
 \left(  1+  \prod_{\alpha} \sp_i^{(\alpha)}
        \right)
\eqno(19c)
$$
It is worth stressing that the above expression
 differs from the partition function of a collection of $n$ non-interacting
 Ising systems {\it without disorder} only because of  the factor
$\prod_{i} (1+  \prod_{\alpha} \sp_i^{(\alpha)})/2$.
Such a term introduces an `effective' interaction between replicas:
 a configuration contributes
 to the annealed partition function $\avd {\Z^n}$ only if
 $\prod_{\alpha} \sp_i^{(\alpha)}=1$
on each site of the dual lattice (plaquette of the original lattice).

Now we can use the mean field approximation to estimate (19), by
introducing the magnetizations
$$
m_{\alpha}=\lim_{N \to \infty} {1\over N} \sum_i^{N} \sp_i^{(\alpha)}
 \quad \alpha=1,\cdots,n
\eqno(20)
$$
Indeed, if we neglect the fluctuations,  the quadratic
  term of (19b) can be estimated as $ \sp_i^{(\alpha)} \, \sp_j^{(\alpha)}
=m_{\alpha}^{2}$  so that
 (16) becomes
$$
\avd{(\Z_N)^n}=
       \sum_{\{\S\} }
e^{N \, 2 \b \sum_{\alpha} m_{\alpha}^{2} }
 \ \prod_{i=1}^{N}  \,
 \left( 1+  \prod_{\alpha} \sp_i^{(\alpha)}
           \right) {1\over 2}
\eqno(21)
$$
The mean field solution can be found by the introduction
 of $n$ auxiliary fieldS
 $\Phi_1,\cdots \Phi_n$. Using the saddle point method,
one has, in the limit $N \to \infty$,
$$
e^{N \, 2 \, \b  \, m_{\alpha}^{2} }\sim  \int_{-\infty}^{\infty}
  d\Phi_{\alpha} \exp \left(
 N \,  2 \, \b \,
 ( 2 \, m_{\alpha} \Phi_{\alpha}  - \Phi_{\alpha}^{2}) \right)
\eqno(22)
$$
As a consequence,  $\avd{(\Z_N)^n}$ is given by the maximum
over $\{\Phi_1,\cdots,\Phi_n\}$  of
$$
 \, \sum_{\{\S\}}
e^{-N \, 2 \b  \sum_{\alpha} \Phi_{\alpha}^2 }
 \ \prod_{i=1}^{N}  \,
{1\over 2}   \left( 1+  \prod_{\alpha} \sp_i^{(\alpha)}
            \right)
       e^{4\b\sum_{\alpha} \Phi_\alpha \sp_i^{(\alpha)}  } \ =
$$
$$
=  \sum_{\{\S\}}
e^{-N \, 2 \b  \sum_{\alpha} \Phi_{\alpha}^2 }
 \ \prod_i^N  \, {1\over 2}  \left[ \
 \prod_{\alpha}^n  e^{4\b \sp_i^{(\alpha)} \Phi_{\alpha} } +
                   \prod_{\alpha}^n  \ \sp_i^{(\alpha)}
       e^{4\b \,  \Phi_\alpha \sp_i^{(\alpha)}  } \ \right ]
\eqno(23)
$$
Now, we can explicitly carry out the sum over
 the $2^{N \, n}$ spin configurations in (23) and obtain
$$
\avd{(Z_N)^n} =\max_{\Phi_1,\cdots,\Phi_n}
e^{-2 \b N \sum_{\alpha=1}^n \Phi_{\alpha}^2 }
  \prod_i^N      {1\over 2}
\left(  \prod_{\alpha}^n  2\cosh(4\b\Phi_{\alpha}) \ + \
\prod_{\alpha}^n  2\sinh(4\b\Phi_{\alpha}  ) \right)
\eqno(24)
$$
In $2d$, it is commonly believed that there is no glass transition
 and no replica symmetry breaking,   so that we expect that the
  maximum of (24) is realized  at  the same $\Phi_{\alpha}=\Phi^*$
 for all  the replicas.
As a consequence, using the replica trick  (18),
 the quenched free energy
 in the mean field approximation reads
$$
 \f (\b)=  -\b^{-1} \,
 \max_{\phi} \left( \,  {1 \over 2} \ln(2 \sinh(8 \, \b \Phi) \, )
 \,  -2 \b \, \Phi^{2} \right)
\eqno(24)
$$
where the maximum of (24) is realized by the value $\Phi^*$,
solution of the self-consistency equation
$$
\coth(8 \b \Phi)=\Phi.
\eqno(25)
$$
The graphical solution of this implicit equation is showed
 in fig 1. One sees that $\Phi^*$ should be
always larger than unity
and at $\b \to \infty$ (infinite temperature $T=\beta^{-1}$ limit)
$\Phi^*=1$. It can appear rather odd that
 in the dual model  the magnetization $\Phi^* \ge 1$.
 This stems from the fact that
 the Gibbs probability measure $\exp(-\b H)$ is
 a signed measure
 because the random coupling is transformed into
 a complex random magnetic field in (17).
 From fig 1, it is also clear that the
 mean field solution does not exhibit  phase transitions
at finite temperature. However, there is an
 essential singularity at $T=0$, since inserting (24) into (15) and (14)
 one sees that $f \sim \exp(1/T)$ for $T \to 0$.

It is possible to explicitly
 solve the self-consistency equation when $\b \to 0$
since (25) becomes
$$
\Phi^* = \left( 8 \,  \b \right)^{-1/2} \, \left(
 1+4\b/3 +O(\b^2) \right)
\eqno(26)
 $$
The zero temperature energy of the mean field solution
is $E_0=-1.5$ while
the numerical simulations [7] give $E_0=-1.404 \pm 0.002$.
In fig 2, we show the free energy as a function
 of $T$.
 One sees that entropy is negative at low temperature,
  thus indicating that the solution is unphysical.
 As a consequence, a better estimate of the ground state energy
 is given   by the maximum of $f(\beta)$, following a standard argument of
 Toulouse and Vannimenus [8],
and one has $E_0 \ge \max_{\beta} f(\beta)=-1.468$.
\bigskip
{\bf 3. BETHE-PEIERLS APPROXIMATION}
\medskip
\noindent
The mean field approximation neglects the short range order,
that can be taken into account by the so-called Bethe-Peierls approximation
 [9-10].
It is still useful to consider the model on the dual lattice
and, moreover, it is convenient
 to work on the internal energy
$$
U(\beta)= {\partial \over \partial \beta} (\beta \, f)
\eqno(27)
$$instead of the free energy.
{}From (14) and (15) one thus has
$$
U(\beta)= -2 \, \coth(2 \beta) -{1\over \sinh(2\beta)}\U(\b)
\eqno(28)
$$
with
$$
\U(\b)={\partial \over \partial \b} \, (\b \f(\b) \,)
\equiv \lim_{n \to 0} \U_n(\b)
$$
where $\f(\b)$ is given by (18)  so that the internal energy of
 $n$ replicas is:
$$
U_n(\b)= \left(\avd{(Z_N)^n} \right)^{-1}
 \sum_{ \{ \S \} }
\sp_k^{(\gamma)} \sp_l^{(\gamma)}
  \prod_i^N {1\over 2}
 \left( 1+ \prod_\alpha^n \sp_i^{(\alpha)} \right)
 \prod_{\alpha}^n e^{\b \sum_{(i,j)} \sp_i^{(\alpha)} \sp_j^{(\alpha)} }
\eqno(29)
$$
that in the limit $n \to 0$ gives $\U(\b)$.

Now comes the key step.
 Let us sum over all the spin couples but the $4n$ nearest neighbor
spins $\sp_1^{(\alpha)}$, $\sp_2^{(\alpha)}$, $\sp_3^{(\alpha)}$,
 $\sp_4^{(\alpha)}$ around the $n$ spin $\sp_0^{(\alpha)}$
 in the numerator of (29).
In order to simplify the notation we shell indicate in the following
all  these $5n$ spins as $\sc$ and the $4n$ lateral ones as $\sq$

       Noting  that
the prefactor $\avd{(Z_N)^n}$ is a constant depending on $n$,
 the expression (29) becomes:
$$
\U_n(\b)=
         \left(\avd{(Z_N)^n} \right)^{-1}
 \sum_{ \{ \sc \} }
\sp_k^{(\gamma)} \sp_l^{(\gamma)}    \,
\Psi_n(\sq) \,
  \left({1\over 2}\right)^5
\prod_{i=0}^4 \left( 1+ \prod_{\alpha}^n \sp_i^{(\alpha)} \right)
\prod_{i=1}^4
 \prod_{\alpha}^n e^{\b \sp_0^{(\alpha)} \sp_i^{(\alpha)} }
\eqno(30)
$$
where $\Psi_n$ is a function of the $4n$ lateral spins $\sq$.
In practice, we are considering
 the $5n$ free spins $\sc$ on replicated crosses which are
 interacting with the mean field generated by the
 other $n(N-5)$ ones.

The ansatz of the Bethe-Peierls approximation consists
in assuming that any function $\Psi_n$ of the spin configurations
such as (30)  might be expressed as
$$
\Psi_n=
\left(\prod_{\alpha=1}^n \prod_{i=1}^4 e^{ \mu \sp_i^{(\alpha)}  }
\right)
  { \avd{(Z_N)^n} \over  z(n,\b,\mu)   }
\eqno(31)
$$
where we have introduce the normalization factor
$$
z(n,\b,\mu)=\sum_{\{\sc\}} W_n(\sc)
\eqno(32)
$$
related to the weight  of the $\sc$ configurations
$$
W_n(\sc)=
 \left({1\over 2}\right)^5
\left( 1+ \prod_{\alpha}^n \sp_0^{(\alpha)} \right)
\prod_{i=1}^4 \left( 1+ \prod_{\alpha}^n \sp_i^{(\alpha)} \right)
 \prod_{\alpha}^n e^{(\b \, \sp_0^{(\alpha)} +\mu)\sp_i^{(\alpha)} }
\eqno(33)
$$
The parameter $\mu$
is a sort of chemical potential  representing the energy  cost
necessary to flip the lateral spins in the opposite direction of $\sp_0$,
 destroying the short range order.
In fact, the Bethe-Peierls approximation is also indicated as
 {\it quasi-chemical} approximation.

As a consequence, the internal energy becomes
$$\U_n(\b)= \, - 2
 \avt{\sp_0 \, \sp_1}_n
\eqno(34)
$$
Here and in the following
 $\avt{{A}}_n$
indicates  the average  of an observable $A$,
$$
\avt{A}_n\equiv z^{-1}(n, \b, \mu) \sum_{\{\sc \}} A \
 W_n(\sc)
\eqno(35)
$$
over the $2^{5n}$ configurations of the spins on the replicated crosses
weighted by $W_n$.
The chemical potential $\mu$ depends on the replica number $n$
 and can be determined through a
 self-consistency equation  given by the requirements
that the average value of the dual spin is invariant under translations,
 i.e.
$$
\avt{ \sp_0 }_n
= \, \avt{\sp_i}_n
\qquad i=1,\cdots,4 \, ; \ \alpha=1,\cdots,n
\eqno(36)
$$
One is interested in the limit $n \to 0$, as usual.
In order to write the self-consistency equation in a simpler form,
it is convenient to introduce the generating function
$$
\phi_n(h,\mu,\b)=
 \ln \sum_{\{\sc\}}  W_n(\sc) e^{h\sum_{\alpha} \sp_0^{(\alpha)} }
\eqno(37)
$$
so that (36) corresponds to require
 $$
\left. {\partial \phi \over \partial h}\right|_{h=0}
= \left. {1\over 4} {\partial \phi\over \partial \mu}\right|_{h=0}
\eqno(38)
$$
where $\phi$ is the quenched generating function,
$$
\phi(h,\mu,\b)=\lim_{n \to 0} {\phi_n \over n}.
$$
The solution of this implicit equation  gives  the value of the chemical
potential $ \mu^*(\b)$ as function of the temperature.
 The internal energy can then be expressed in terms of
 the generating function as
$$
\U(\b)= \, -{1\over 2}
 \left. {\partial \phi \over \partial \b}\right|_{h=0
 \, , \,  \mu^*(\b)}
\eqno(39)
$$
In order to obtain the quenched generating function $\phi$, we should
perform some algebraic manipulations.
After performing the sum over the $2^{4nN}$ configurations
 $\{\sq\}$ in (37) we remain with a sum over the configurations
$\Sz \equiv \sp_0^{(1)},\cdots,\sp_0^{(n)}$
$$
2^5 \, e^{\phi_n}
=
 \sum_{\{\Sz\}}
 \left( 1+\prod_{\alpha} \sp_0^{(\alpha)} \right)
 \left(
\prod_{\alpha} 2 \cosh \eta^{(\alpha)}
+
\prod_{\alpha} 2 \sinh\eta^{(\alpha)}
 \right)^4
 e^{h \sum_{\alpha} \sp_0^{(\alpha)} }
$$
$$
=
 \sum_{\{\Sz\}}
\sum_{k=0}^4 {4 \choose k}
  \left( 1+\prod_{\alpha} \sp_0^{(\alpha)} \right)
   \prod_{\alpha}
 \left(2 \cosh\eta^{(\alpha)} \right)^{4-k}
  \prod_{\alpha} \left( 2 \sinh\eta^{(\alpha)}
 \right)^k
 e^{h  \sp_0^{(\alpha)} }
$$
$$
=
 \sum_{\{\Sz\}}  \sum_{k=0}^4 \sum_{j=\pm1}
 {4 \choose k}
  \ \prod_{\alpha}  \ (\sp_0^{(\alpha)})^{{1+j\over 2}}
   \left(2 \cosh\eta^{(\alpha)}  \right)^{4-k}
 \,  \left( 2 \sinh\eta^{(\alpha)} \right)^k
 e^{h \sp_0^{(\alpha)} }
\eqno (40)
$$
where we have introduced the variable
$$
\eta^{(\alpha)}\equiv  \mu+ \b \, \sp_0^{(\alpha)}
\eqno(41)
$$
 for simplifying the notation.
Now,  the previous sum   has been
obtained by an annealed average over the disorder, i.e.
$$\sum_{k=0}^4 \sum_{j=\pm 1} 2^{-5}
{4 \choose k} A_{k,j}=\avd{A}
\eqno(42)
$$
where one has
$$
A_{k,j}=
 \sum_{\{\Sz\}}
  \prod_{\alpha} (\sp_0^{(\alpha)})^{{1+j\over 2}}
    \left(2 \cosh \eta^{(\alpha)} \right)^{4-k}
 \,  \left( 2 \sinh\eta^{(\alpha)} \right)^k
 e^{h \sp_0^{(\alpha)} }
$$
$$
=
\left(  \sum_{\{\sp_0\}}
   \sp_0^{{1+j\over 2}}
    \left(2 \cosh\eta \right)^{4-k}
 \,  \left( 2 \sinh\eta \right)^k
 e^{h \sp_0}
\right)^n
\equiv(a_{k,j})^n
\eqno(43)
$$
with $\eta=\mu+\beta \sp_0$. Noting that
$$
\lim_{n \to 0} {1\over n} \ln \avd{a^n}=\avd{\ln a}
$$
eventually one can write the quenched generating function as
$$
\phi=
{1\over 2^5} \sum_{j=\pm 1} \sum_{k=0}^4
 {4 \choose k}  \, \ln \sum_{\sp_0=\pm 1}
  \, e^{h\sp_0}   \, \sp_0^{(1+j)/2} \
 \cosh^{4-k}\eta \,  \sinh^{k}\eta
\eqno(44)
$$
Here and in the following we omit to write the constant additive term
$4 \ln2$ in $\phi$. Note that such a term disappears in the derivatives.
The first sum over $j$  in (44) can be performed by a trick.
 Let us use  an auxiliary spin ${\sp'}_0=\pm 1$ with equal probability
 so that
$$
\phi=
{1\over 2^{5}} \sum_{k=0}^4
 {4 \choose k}   \ln
 \sum_{\sp_0\, , \, \sp_0'=\pm 1}
   e^{h(\sp_0+\sp_0')}   \, \sp_0
 \left(\cosh\eta \,  \cosh\eta' \right)^{4-k}
\left(\sinh\eta \,  \sinh\eta' \right)^{k}
\eqno(45)
$$
with $\eta'\equiv \mu+\b \sp_0'$.
It is easy  to realize that
   if $\sp_0$ and  $\sp_0'$ have opposite sign, the
 contribution to the sum vanishes.
We can limit ourselves to consider the case of equal sign, so that
(45) becomes
$$
\phi=
{1\over 2^{5}} \sum_{k=0}^4
 {4 \choose k}   \ln
 \sum_{\sp_0=\pm 1} \sp_0 \,
  \, e^{2h \sp_0} \,
 \cosh^{2(4-k)}\eta \,  \sinh^{2k}\eta
\eqno(46)
$$
Moreover in the limit $h \to 0$, one has
$$
\sp_0 \,  e^{2h \sp_0}  \sim \sp_0 \, + \, 2 \, h
\eqno(47)
$$
It follows that $\avt{\sp_0}$ is
$$
\left. {\partial \phi \over \partial h}\right|_{h=0}
=
{1\over 2^5} \sum_{k=0}^4
 {4 \choose k} S_k^{-1} \
\sum_{\sp_0=\pm 1} 2 \cosh^{2(4-k)}\eta \,  \sinh^{2k}\eta
\eqno(48)
$$
with
$$
S_k=
 \sum_{\sp_0=\pm 1} \sp_0 \,
 \cosh^{2(4-k)}\eta \,  \sinh^{2k} \eta
\eqno(49)
$$
while $\sum_{i=1}^4 \avt{ \sp_i} $ is
$$
\left. {\partial \phi \over \partial \mu}\right|_{h=0} =
{1\over 2^5} \sum_{k=0}^4
 {4 \choose k}  S_k^{-1}
\sum_{\sp_0=\pm 1} \sp_0 Y_k(\eta)
\eqno(50)
$$
with
$$
Y_k(\eta)=   \cosh^{2(4-k)}\eta \,  \sinh^{2k}\eta
 \,  \left[
2(4-k) \tanh\eta \, + \, 2k \, \coth \eta \, \right]
\eqno(51)
$$
Inserting (45) and (46) into the self-consistency equation (38),
the chemical potential
 $\mu^*$ can be obtained as function of the dual inverse temperature $\b$.
Once determined the value $\mu^*$ , the internal energy  is given by
a derivative of the generating function. In particular one has
$$
\U(\b)=-{1\over 2} \left.
{\partial \phi \over \partial \b}\right|_{h=0 \, , \,
  \mu^*(\b)}=
- {1\over 2^6} \sum_{k=0}^4
 {4 \choose k}  S_k^{-1} \
\sum_{\sp_0=\pm1}  Y_k(\eta)
\eqno(52)
$$
In fig 3 we show $\mu^*/\b$ as a function of $\b$, where one observes
 that for $\b \to \infty$
(limit of high temperature $T$ of the original lattice),
 $\mu^*/\b  \to 3$. This can be understood noting that
 each one of the lateral free spins of the cross interacts
 with three other spins so that, at zero dual temperature $\b^{-1}$,
the energy lost in a flip is exactly equal to $3$.
Fig 4  illustrates the graphical solution of (38) by plotting
$$
\lim_{n \to 0} \ \left(
\avt{\sp_0}_n \, - \, {1\over 4} \sum_{i=1}^4 \avt{\sp_i}_n  \right)
\eqno(53)
$$
as function of $\mu $ at three different temperatures.
 The solution $\mu^*$ of the self-consistency
equation  is given by the intersection of the function
 with the horizontal axes. We look only for
 real solutions.  At large $\b$,  there exists only one
solution. However, fig 4 shows that at low $\b$ two solutions appear,
 and for $\b<0.031$ there is no real positive solution.
 The internal energy $U$ given by the Bethe-Peierls approximation
 together with the annealed energy $U_a=-2\tanh(\beta)$
 and the mean field result are
 plotted as function of the temperature in fig 5.
 One sees that for $\b<0.05$, i.e. $T<0.667..$,  the energy
increases at lowering the temperature
  while the chemical potential $\mu^*$ decreases,
 indicating that the Bethe-Peierls solution becomes unphysical.
 The ground state energy can be estimated
 by the minimum value assumed by the internal energy,  i.e.
$$
E_0 \approx \min_{\b} U(\b) = U(\b=0.05..)=-1.3975
$$
It is extremely close to the numerical estimate of [7]
  $E_0=-1.404 \pm 0.002$.

Finally, we want to mention that it remains open the problem to understand
 whether,  with an appropriate  ansatz of replica symmetry breaking,
 one can obtain a solution of the self-consistency equation
  for $\b <0.03$. It is indeed well known that mean field approach
 can give phase transitions, even when they are absent in
 the original model.
\bigskip
{\bf 4. IMPROVED BETHE-PEIERLS APPROXIMATION}
\medskip
\noindent
In order to improve
 the Bethe-Peierls approximation
 in the framework of the replica method,
we introduce a second variational parameter $\gamma$,  beyond
the chemical potential $\mu$,  that puts in
 interaction the different replicas.
 In other terms, we replace the (standard) ansatz (31) with
$$
\Psi_n=
\prod_{\alpha} \prod_{i=1}^4 e^{ \mu \sum_{\alpha}\sp_i^{(\alpha)}
+\gamma \sum_{\alpha>\beta}
\sp_i^{(\alpha)} \sp_i^{(\beta)}  }
  { \avd{(Z_N)^n} \over  g(n,\b,\mu,\gamma)   }
\eqno(54)
$$
where we have introduced the normalization factor
$$
g(n,\b,\mu,\gamma)=\sum_{\{\sc\}} P_n(\sc)
\eqno(55)
$$
related to the weight  of the $\sc$ configurations
$$
P_n(\sc)= W_n e^{\gamma \sum_{\alpha>\beta}
\sp_i^{(\alpha)} \sp_i^{(\beta)}  }
\eqno(56)
$$
with $W_n$ given by (33).
 We shall indicates the average of an observable $A$ over this new
 normalized weight by $<<A>>_n$.

The two variational parameters are determined by the couple
self-consistency equations obtained in the limit $n \to 0$
by
$$
<<{\sp_0}>>_n=<<{\sp_i}>>_n
\eqno(57a)
$$
$$
<<{\sp_0}^{(\alpha)} \, {\sp_0}^{(\beta)}>>_n=
<<{\sp_i}^{(\alpha)} \, {\sp_i}^{(\beta)}>>_n
\eqno(57b)
$$
with $i=1,\cdots,4$ and $\alpha,\beta=1,\cdots,n$.

The ansatz  proposed here is related to a hypothesis of
existence of a glassy phase.
In fact,  one can apply this approximation scheme to the the solution of the
 random coupling Ising model directly on the original lattice in
$d$-dimensions [11].
 In this case, after performing the limit $d \to \infty$, one
 obtains the Parisi solution [2] of the Sherrington-Kirkpatrick
model.

Following the same idea of the previous section, let us introduce the
 generating function
$$
\psi_n(h,\ell,\mu,\gamma,\b)=
 \ln \sum_{\{\sc\}}  P_n(\sc) e^{h\sum_{\alpha} \sp_0^{(\alpha)}
+\ell  \sum_{\alpha>\beta}
\sp_i^{(\alpha)} \sp_i^{(\beta)}
 }
\eqno(58)
$$
so that (57a) and (57b) correspond to require
 $$
\left. {\partial \psi \over \partial h}\right|_{h=0,\ell=0}
= \left. {1\over 4} {\partial \psi\over \partial \mu}\right|_{h=0,\ell=0}
\eqno(59a)
$$
and
    $$
\left. {\partial \psi \over \partial \ell}\right|_{h=0,\ell=0}
= \left. {1\over 4} {\partial \psi\over \partial \gamma}\right|_{h=0,\ell=0}
\eqno(59b)
$$
where $\psi$ is the quenched generating function,
$$
\psi(h,\mu,\gamma,\b)=\lim_{n \to 0} {\psi_n \over n}.
\eqno(60)
$$
The solution of this implicit equation  gives  the values
$ \mu^*(\b)$ and $\gamma^*$ as function of the temperature.
 The internal energy can then be expressed in terms of
 the generating function as
$$
\U(\b)= \, -{1\over 2}
 \left. {\partial \psi \over \partial \b}\right|_{h=0,\ell=0
 \, , \,  \mu^*(\b),\gamma^*(\b)}
\eqno(61)
$$

Let us now simplify as much as possible the self-consistency equations.
It is convenient to use the standard gaussian identity
$$ \exp\left( \ell
\sum_{\alpha>\beta}
\sp_0^{(\alpha)} \sp_0^{(\beta)}
 \right)=
\exp\left( {\ell\over 2} \, \left(\sum_{\alpha} \sp_0^{(\alpha)}\right)^2
-{\ell n\over 2} \right)=
$$
$$
=
\int {dx_0\over \sqrt{2\pi}}
\exp\left(-{\ell n\over 2 }+
 \sqrt{\ell} x_0 \sum_{\alpha} \sp_0^{(\alpha)}-{x_0^2\over 2}
\right)
\eqno(62)
$$
in order to write the generating function as
$$
\psi_n = -{(4\gamma +\ell) \, n\over 2} +
\ln \left( \, 2^{-5} \, \int \prod_{i=0}^4
  {dx_i\over \sqrt{2\pi}} e^{-{x_i^2\over 2}} \right.
$$
$$
 \left. \sum_{\{\Sz\}}
 \left( 1+\prod_{\alpha} \sp_0^{(\alpha)} \right)
 \left(
\prod_{\alpha} 2 \cosh\omega_i^{(\alpha)}
+
\prod_{\alpha} 2 \sinh\omega_i^{(\alpha)}
 \right)
 e^{(h+x_0\sqrt{\ell}) \sum_{\alpha} \sp_0^{(\alpha)} }
\right)
\eqno(63)
$$
with
$$
\omega_i^{(\alpha)}=\eta^{(\alpha)}+x_i \sqrt{\gamma}
\eqno(64)
$$
Using the same  `algebraic' strategy  that in the previous section
 leads to (44),
we obtain the quenched generating function as
$$
\psi= \, -2\gamma \, - \, {\ell\over 2} \, +
{1\over 2^5}
\int \prod_{i=0}^4  {dx_i\over \sqrt{2\pi}} e^{-{x_i^2\over 2}}   \,
 \sum_{j_1,j_2,j_3,j_4=\pm 1}
$$
$$
\ln \left(\sum_{\sp_0=\pm 1}
 \sp_0
  \, e^{2 \, (h+\sqrt{\ell}x_0) \sp_0}   \,
\prod_{i=1}^4
(\cosh \omega_i)^{1+j_i} \,
(\sinh \omega_i)^{1-j_i}  \right)
\eqno(65)
$$
with
$$
\omega_i=\eta+x_i\sqrt{\gamma}=\b \, \sp_0+\mu+x_i \sqrt{\gamma}.
$$
The constant additive term $4 \ln 2$ is again omitted.

By derivating $\psi$ one has the self-consistency equations   (57)
for $\mu^*$ and $\gamma^*$   in terms of the sum of
$5$ gaussian integrals.
A careful analytic and numerical study of these equations
might enlighten the nature
of spin glasses in low dimensional systems.
 \bigskip
{\bf 5. CONCLUSIONS}
\medskip
\noindent
Let us briefly summarize our main results:

\item{ (1)} formulation of the random coupling Ising model on the dual
 lattice made of square plaquettes. The dual model has signed
 Gibbs probability measure as the random coupling is transformed into a
 random complex magnetic field.

 \item{(2)} Application of the mean field approximation to
 the two dimensional dual model in the framework of the replica method.
 We find the solution using a replica symmetry ansatz, obtaining
 a  good estimate of the ground state energy:
 $E_0=-1.468$, compared with  the numerical result of [7]
 $E_0=-1.404 \pm 0.002$.

\item{(3)} Application of the Bethe-Peierls approximation.
 It  gives a very accurate estimate of the ground state
 energy ($E_0=-1.3975$) although it becomes unphysical below $\b=0.05$
 and there is no real solution of the self-consistency
 equation for $\b<0.03$.

\item{(4)}
   Improvement scheme of the Bethe-Peierls approximation
    by considering a second variational parameter that puts in interaction
     different replicas of the dual model.

There are still many problems that remain open in our approach.
   As major issue,  it would be interesting to understand
 whether  the ansatz proposed in section 4, or other similar  assumptions
  lead to a solution of the self-consistency equations
 of the Bethe-Peierls approximation at low $\b$,
 that allow one  to decide whether  a transition
to a glassy phase is present at low dimension.

  The dual transformation is indeed a very powerful tool
 for determining the critical temperature in non-disordered  systems
and our method  might give some results in this direction.
Least but not last, we plan to find a cluster expansion scheme
that permits to improve the mean field in a systematic way.
\bigskip
{\bf ACKNOWLEDGEMENTS}
\medskip
\noindent
We acknowledge the financial support
 ({\it Iniziativa Specifica} FI3) of the I.N.F.N.,
  National Laboratories  of Gran Sasso.
 GP is grateful to Nordita and MS to the
 Institut Sup\'erieure Polytechnique
 de Madagascar for hospitality.
 We thank S. Franz, J. Raboanary and  S. Solla
 for useful comments and criticisms.
\vfill\eject
\noindent
{\bf References}
\bigskip
\bigskip
\item{[1]}
  M. Serva, G. Paladin and J. Raboanary,
Cond-Mat  preprint n 9509005 of the xxx.lanl.gov e-print archive (1995)
\bigskip
\item{ [2]}
M. Mezard, G. Parisi and M. Virasoro, {\it Spin glass theory and beyond},
 World Scientific Singapore 1988
\bigskip
\item{ [3]}
G. Toulouse, Commun. Phys.  {\bf 2}, 115 (1977)
\bigskip
\item{ [4]}
H. A. Kramers and G. H. Wannier,
Phys. Rev. {\bf 60}, 252 (1941)
\bigskip
\item{[5]}
G. G. Batrouni,
Nucl. Phys. B {\bf 208}, 12 (1982);
G. G. Batrouni and M.B. Halpern,
Phys. Rev. D {\bf 30}, 1782 (1984)
\bigskip
\item{[6]}
G. G. Batrouni, E. Dagotto,  and A. Moreo,
Phys. Lett. {\bf 155B}, 263 (1985)
\bigskip
\item{[7]}
L. Saul and M. Kardar,
Phys. Rev. E {\bf 48}, 48 (1993)
\bigskip
\item{ [8]}
G. Toulouse and J. Vannimenus,
Phys. Rep. {\bf 67}, 47 (1980)
\bigskip
\item{ [9]}
H. Bethe
Proc. R. Soc. London {\bf 150}, 552 (1935)
\bigskip
\item{ [10]}
R. Peierls
Proc. R. Soc. London {\bf 154}, 207 (1936)
\bigskip
\item{[11]}
  M. Serva, unpublished
\vfill\eject
\vskip 0.8truecm
\centerline {\bf Figure Captions}
\vskip 0.5truecm
\noindent

\item{Fig. 1}
Graphical solution of the implicit equation (25)
 at $T=\beta^{-1}=1$  corresponding to  $\b=0.136..$.
 The full lines are $\coth(8 \b \Phi)$ versus $\Phi$
 and the straight line $\Phi=\Phi$.

\bigskip
\item{Fig. 2}
Annealed free energy $f_a$ given by (4) (dashed line) and the
mean field solution (full line)  versus temperature $T=\beta^{-1}$.
The dashed lines are the Maxwell constructions obtained by imposing
that the free energy is a monotonous non-decreasing function of $T$.
 The annealed solution estimates a ground state energy
$E_0\ge-1.559$; the mean field solution gives
 $E_0\ge -1.468$; the numerical result of [5] is $E_0=-1.404\pm 0.002$

\bigskip
\item{Fig. 3}
Chemical potential $\mu^*/\b$ as function of the dual inverse temperature $\b$.

\bigskip
\item{Fig. 4}
The Graphical solution of the self-consistency equation (36)
 is given by the intersection of (53) with the horizontal axes.
We illustrate three different cases: $\b=0.1$ (full line),
 $\b=0.05$ (dashed line) and $\b=0.028$ (dotted line).

\bigskip
\item{Fig. 5}
Internal energy given by the annealed approximation,
 i.e. $U_a=-2 \tanh \beta$
  (dashed line), by the mean field approximation (dash-dotted line)
  and by the Bethe-Peierls solution (full line) $U(T)$
  versus temperature $T=\beta^{-1}$.
The dotted line is obtained by imposing
that the Bethe-Peierls
 internal energy is a monotonous non-increasing function of $T$.
\bye